\documentclass[useAMS,usenatbib]{mn2e}

\usepackage{enumerate}
\usepackage{gensymb}
\usepackage{ textcomp }
\usepackage{graphicx}
\usepackage[caption=false]{subfig}

\title[Modelling of the Spectral Energy Distribution of the Radio Galaxy Fornax~A]{Modelling of the Spectral Energy Distribution of Fornax~A: Leptonic and Hadronic Production of High Energy Emission from the Radio Lobes}

\author[B. McKinley et al.]
{B.~McKinley,$^{1,2}$\thanks{E-mail:benjamin.mckinley@anu.edu.au}
R.~Yang,$^{3}$
M.~L\'{o}pez-Caniego,$^{4}$
F.~Briggs,$^{1,2}$
N.~Hurley-Walker,$^{5}$
\newauthor
R.~B.~Wayth,$^{5,2}$
A.~R.~Offringa,$^{1,2}$
R.~Crocker,$^{1}$
G.~Bernardi,$^{6,7,8}$
P.~Procopio,$^{9,2}$
\newauthor
B.~M.~Gaensler,$^{10,2}$
S.~J.~Tingay,$^{5,2}$
M.~Johnston-Hollitt,$^{11}$
M.~McDonald,$^{12}$
\newauthor
M.~Bell,$^{13,2}$
N.~D.~R.~Bhat,$^{5,2}$
J.~D.~Bowman,$^{14}$
R.~J.~Cappallo,$^{15}$
B.~E.~Corey,$^{15}$
\newauthor
A.~A.~Deshpande,$^{16}$
D.~Emrich,$^5$ 
A.~Ewall-Wice,$^{12}$
L.~Feng,$^{12}$
R.~Goeke,$^{12}$
\newauthor
L.~J.~Greenhill,$^7$
B.~J.~Hazelton,$^{17}$
J.~N.~Hewitt,$^{12}$ 
L.~Hindson,$^{11}$
D.~Jacobs,$^{14}$
\newauthor
D.~L.~Kaplan,$^{18}$
J.~C.~Kasper,$^{7,19}$
E.~Kratzenberg,$^{15}$
N.~Kudryavtseva,$^{5}$
\newauthor
E.~Lenc,$^{10,2}$
C.~J.~Lonsdale,$^{15}$
M.~J.~Lynch,$^5$
S.~R.~McWhirter,$^{15}$
D.~A.~Mitchell,$^{9,2,13}$
\newauthor
M.~F.~Morales,$^{17}$
E.~Morgan,$^{12}$
D.~Oberoi,$^{20}$
S.~M.~Ord,$^{5,2}$
B.~Pindor,$^{9,2}$
T.~Prabu,$^{16}$
\newauthor
J.~Riding,$^{9,2}$
A.~E.~E.~Rogers,$^{15}$
D.~A.~Roshi,$^{21}$\thanks{The National Radio Astronomy Observatory is a facility of the National Science Foundation operated under cooperative agreement by Associated Universities, Inc.}
N.~Udaya~Shankar,$^{16}$
\newauthor
K.~S.~Srivani,$^{16}$
R.~Subrahmanyan,$^{16,2}$
M.~Waterson,$^{5,1}$
R.~L.~Webster,$^{9,2}$
\newauthor
A.~R.~Whitney,$^{15}$
A.~Williams,$^{5}$
C.~L.~Williams$^{12}$
\\
$^{1}$Research School of Astronomy and Astrophysics, Australian National University, Canberra, ACT 2611, Australia\\
$^{2}$ARC Centre of Excellence for All-sky Astrophysics (CAASTRO), Australian National University, Canberra, ACT 2611, Australia\\
$^{3}$High Energy Astrophysics Theory Group, Max-Planck-Institut f\"{u}r Kernphysik, Heidelberg 69029, Germany\\
$^{4}$Instituto de F\'isica de Cantabria (CSIC-UC), Avda. Los Castros S/N, 39005, Santander, Spain\\
$^{5}$International Centre for Radio Astronomy Research, Curtin University, Bentley, WA 6102, Australia\\
$^{6}$Square Kilometre Array South Africa (SKA SA), Pinelands 7405, South Africa\\
$^{7}$Harvard-Smithsonian Center for Astrophysics, Cambridge, MA 02138, USA\\
$^{8}$Department of Physics and Electronics, Rhodes University, Grahamstown 6140, South Africa\\
$^{9}$School of Physics, The University of Melbourne, Parkville, VIC 3010, Australia\\
$^{10}$Sydney Institute for Astronomy, School of Physics, The University of Sydney, NSW 2006, Australia\\
$^{11}$School of Chemical \& Physical Sciences, Victoria University of Wellington, Wellington 6140, New Zealand\\
$^{12}$Kavli Institute for Astrophysics and Space Research, Massachusetts Institute of Technology, Cambridge, MA 02139, USA\\
$^{13}$CSIRO Astronomy and Space Science, Marsfield, NSW 2122, Australia\\
$^{14}$School of Earth and Space Exploration, Arizona State University, Tempe, AZ 8528USA\\
$^{15}$MIT Haystack Observatory, Westford, MA 01886, USA\\
$^{16}$Raman Research Institute, Bangalore 560080, India\\
$^{17}$Department of Physics, University of Washington, Seattle, WA 98195, USA\\
$^{18}$Department of Physics, University of Wisconsin--Milwaukee, Milwaukee, WI 53201, USA\\
$^{19}$Department of Atmospheric, Oceanic and Space Sciences, University of Michigan, Ann Arbor, MI 48109, USA\\
$^{20}$National Centre for Radio Astrophysics, Tata Institute for Fundamental Research, Pune 411007, India\\
$^{21}$National Radio Astronomy Observatory, Charlottesville and Greenbank, USA\\
}

\begin{document}


\pagerange{\pageref{firstpage}--\pageref{lastpage}} \pubyear{2014}

\maketitle

\label{firstpage}
\clearpage

\begin{abstract}

We present new low-frequency observations of the nearby radio galaxy Fornax~A at 154~MHz with the Murchison Widefield Array, microwave flux-density measurements obtained from {\it WMAP} and {\it Planck} data, and $\gamma$-ray flux densities obtained from \emph{Fermi} data. We also compile a comprehensive list of previously published images and flux-density measurements at radio, microwave and X-ray energies. A detailed analysis of the spectrum of Fornax~A between 154~MHz and 1510~MHz reveals that both radio lobes have a similar spatially-averaged spectral index, and that there exists a steep-spectrum bridge of diffuse emission between the lobes. Taking the spectral index of both lobes to be the same, we model the spectral energy distribution of Fornax~A across an energy range spanning eighteen orders of magnitude, to investigate the origin of the X-ray and $\gamma$-ray emission. A standard leptonic model for the production of both the X-rays and $\gamma$-rays by inverse-Compton scattering does not fit the multi-wavelength observations. Our results best support a scenario where the X-rays are produced by inverse-Compton scattering and the $\gamma$-rays are produced primarily by hadronic processes confined to the filamentary structures of the Fornax~A lobes. \\

\end{abstract}

\begin{keywords}
galaxies: individual (NGC1316) - galaxies: active - radio continuum: galaxies 
\end{keywords}

\section{Introduction}

A number of radio galaxies have been shown to have associated $\gamma$-ray emission \citep{2fgl}. In the case of the nearest radio galaxy, Centaurus~A, \emph{Fermi} Large Area Telescope (\emph{Fermi}-LAT; \citealt{LAT}) observations revealed that the $\gamma$-rays originate predominantly in the extended radio lobes \citep{abdo}, rather than the central active galactic nucleus (AGN). The mechanism for producing these $\gamma$-rays was interpreted by \citet{abdo} as inverse-Compton (IC) radiation from the up-scattering of cosmic microwave background (CMB) and extragalactic background light (EBL) photons by relativistic electrons. Using a larger {\it Fermi}-LAT data set, \citet{yang} found that a contribution from proton interactions in the lobe plasma could also account for the shape of the spectral energy distribution (SED) of the Centaurus A lobes. Further observational and theoretical investigation into the origin of the $\gamma$-ray emission from the lobes of radio galaxies is required to determine the relative contributions of the physical mechanisms responsible.

In the IC scattering model of $\gamma$-ray production, we assume that a single population of highly relativistic electrons, with a power-law energy distribution, is responsible for up-scattering seed photons (the most abundant of which are CMB photons) to higher energies. The power-law energy distribution of the electrons results in a power-law spectrum of the synchrotron emission such that the flux density, $S_{\nu}$, is given by $S_{\nu}\propto \nu^{\alpha_r}$, where $\nu$ is frequency and $\alpha_r$ is the radio spectral index. The power-law energy distribution of electrons also results in an IC emission spectrum at higher energies that has the same spectral shape as the synchrotron spectrum. The electron energy index, $p$, which characterizes the shape of the underlying electron energy distribution, is given by $p=2\alpha_r+1$. 

In the hadronic scenario for the production of $\gamma$-rays in radio galaxy lobes, nonthermal cosmic-ray protons produce mesons through proton-proton (p-p) collisions and $\gamma$-rays result from the decay of the neutral pion component. This mechanism has been proposed as an explanation for the $\gamma$-ray emission observed in the so-called {\it Fermi} bubbles of our own Galaxy \citep{crocker1,crocker2}.

Fornax~A was the first radio galaxy shown to emit X-ray IC radiation resulting from the up-scattering of CMB photons by relativistic, synchrotron-emitting electrons in its lobes \citep{laurent,feigelson} and has also been detected as a $\gamma$-ray point source by {\it Fermi} \citep{2fgl}.The host galaxy is the elliptical NGC~1316, which lies at a distance of 18.6~Mpc \citep{madore}. 

In this paper we investigate the origin of the Fornax~A $\gamma$-rays through a detailed analysis of the synchrotron emission in the lobes and broadband SED modelling using a combination of new low-frequency data from the Murchison Widefield Array (MWA; \citealt{lonsdale,tingay,bowman2013}), microwave data from {\it Planck} \citep{planck_sat,planck_results_2013} and {\it WMAP} \citep{bennett2003a,bennett2003b}, $\gamma$-ray data from \emph{Fermi}-LAT and previously published data from a range of ground and space-based telescopes. In Section~2, we discuss previous observations and analyses of the Fornax~A radio lobes. In Section~3, we provide details of the new observations and data reduction procedures and describe the previously published data used. In Section~4, we present a spatially resolved analysis of the low-frequency spectral index of the lobes of Fornax~A. We then fit the multi-wavelength data, using different emission models, to investigate various scenarios for the production of both the X-ray and $\gamma$-ray photons. We discuss the results of our analyses in Section~5.

\section{Previous observations and analyses of Fornax~A}

Fornax~A was one of the earliest identified extragalactic radio sources \citep{mills1954}. The flux density of the entire source has been measured over a wide range of radio frequencies, from 5~MHz \citep{ForA_5MHz} to 5~GHz \citep{5GHz}. However, low-frequency observations (below 400~MHz) lacked the angular resolution required to resolve the two radio lobes and were also unable to resolve the fainter, compact core, which is clearly evident in the higher-frequency images (e.g. \citealt{fomalont}) and in this work. 

\citet{bernardi} examined the radio spectrum of Fornax~A between 5 and 1415~MHz using previously published data and measured the total flux density of Fornax~A using the MWA~32-tile prototype. By fitting a power-law spectrum to the measurements between 30 and 400~MHz (excluding two data points at 100~MHz and one point at 400~MHz, which appeared to be affected by systematic errors) they calculated a spectral index of $-0.88\pm0.05$. Their measured total flux density was $519\pm26$ Jy at 189~MHz. The MWA Commissioning Survey (MWACS; Hurley-Walker, private communication) has also measured the flux density of Fornax A with sub-array configurations of the MWA during science commissioning. Since the sub-arrays had insufficient short spacings to fully sample the large-scale structure of Fornax~A, they quote lower limits on the flux densities, which are $>786$, $>668$ and $>514$~Jy at 120, 150 and 180~MHz, respectively.

There have been several confirmations of the X-ray IC emission from the lobes of Fornax~A since the work of \citet{laurent} and \citet{feigelson}. \citet{kaneda} observed IC X-rays in the lobes of Fornax~A and compared their measurements to previously-published radio data to estimate a magnetic field strength in the lobes of 2-4 $\mu$G. In their analysis, they derived a radio spectral index for the entire source of $\alpha_r=-0.9\pm0.2$ from three published data points at 408~MHz \citep{cameron}, 1.4~GHz \citep{ekers} and 2.7~GHz \citep{shimmins}. Subsequently, \citet{isobe}, using X-ray data on the east lobe of Fornax~A from \emph{XMM-Newton}, improved on the IC analysis by using a more comprehensive set of published radio data. They obtained a radio spectral index, again using data for the entire source, of $\alpha_r=-0.68\pm0.10$ and used this value to derive the physical quantities of the lobes. 

\citet{tashiro} performed an updated IC analysis, this time on the west lobe and going to a higher X-ray energy of 20~keV. The higher X-ray energy means that the same population of electrons (in terms of energy) is being sampled in both the synchrotron (radio) and IC (X-ray) measurements. \citet{tashiro} also fixed the radio spectral index at $-0.68$, as derived by \citet{isobe}, when fitting the SED and deriving the physical quantities of the lobes. \citet{isobe} and \citet{tashiro} also report best-fit X-ray spectral indices of $-0.62^{+0.15}_{-0.24}$ \citep{isobe} and $-0.81\pm0.22$ \citep{tashiro} for the east and west lobes, respectively. The X-ray spectral indices for both lobes are consistent with the radio spectral index of $-0.68\pm0.10$, as is expected in the IC emission model.

In all of the above cases, the available radio data used did not allow evaluation of the spectral index for each of the Fornax~A lobes individually and it was assumed that the spectral behaviour of both lobes was the same. Since the errors on the X-ray spectral indices are large, it is conceivable that the lobes have different electron energy indices. For accurate modelling of the broadband SED it is therefore important to determine if the radio spectral index (and therefore the electron energy index) of each of the lobes is indeed similar.

\section{Multi-wavelength Data}

\subsection{154~MHz MWA observations and data reduction}

The MWA \citep{lonsdale,tingay,bowman2013} is a new low-frequency radio interferometer array that began science operations in 2013 July. It consists of 128 antenna tiles, each containing 16 crossed-dipole antennas above a conducting ground plane. The tiles are pointed electronically using analogue beamformers. The MWA is an official Square Kilometre Array precursor instrument, located at the Murchison Radio Observatory in Western Australia at a latitude of $-26\fdg7$. 

Observations of Fornax~A were made with the MWA on 2013 August 21, comprising of seven snapshot observations, each of 112~s duration and using the same beamformer delay settings. The centre frequency was 154.315~MHz and the full 30.72~MHz bandwidth was used in calibration and imaging. The data were converted from their raw format into {\sc casa} measurement sets using {\sc cotter}, a data conversion pipeline that implements radio-frequency interference (RFI) flagging with {\sc aoflagger}  \citep{offringa2010,offringa2012}. 

An initial set of complex antenna gain solutions was obtained using a single snapshot observation of the bright calibrator source Pictor~A (Pic~A). The VLA image of Pic~A \citep{perley} at 1.4~GHz was re-scaled to match the flux-density value of Pic~A at 154~MHz according to \citet{jacobs} and multiplied by the MWA primary beam, as calculated analytically for the chosen antenna tile pointing. This model was then used to calibrate the data using the {\sc bandpass} task in {\sc casa}. This process does not take into account the intrinsic differences in the diffuse versus compact structure of Pic~A at 1.4~GHz and 154~MHz, which will affect the relative intensities of the response to short and long baselines. However, this effect is considered negligible compared to the inaccuracies resulting from transferring calibration solutions between observations taken with different beamformer settings, which is rectified through self-calibration.   

The initial calibration solutions from Pic~A were transferred to one of the Fornax~A snapshots and an image was made using {\sc wsclean} \citep{wsclean}. The advantage of {\sc wsclean} over other packages is the high speed at which we were able to produce large, widefield images in a standard projection, which could then be used for self-calibration. A 7000 by 7000-pixel image was produced for both of the instrumental polarizations and the resulting clean-component model was placed into the model column of the measurement set. For the self-calibration iterations, {\sc wsclean} was stopped when the first negative component was produced. Self-calibration was then performed in {\sc casa} with the {\sc bandpass} task, in which both the amplitude and phase are solved for. The uv range for calibration was restricted to exclude baselines shorter than 30~wavelengths, since diffuse emission on these scales is not represented in the calibration model.

Three self-calibration iterations were performed and the final set of calibration solutions was then applied to each of the seven Fornax~A snapshots. Both instrumental polarizations were then imaged for each snapshot using {\sc wsclean} with a pixel size of 0.5~arcmin and an image size of 7000 by 7000~pixels. A value of 100 was used for the number of `wlayers' \citep{wsclean} and major iterations were used as per the Cotton-Schwab \citep{schwab} algorithm. The resulting fourteen images were then mosaiced together, using weights for each pixel that were derived from the primary beam shape, which was calculated analytically for each snapshot and each antenna polarization. 

Self-calibration is known to affect the flux-density scale of MWA images. To check the flux-density scale, 16 unresolved point sources from the Culgoora \citep{culgoora,culgoora1995} catalogue were identified within $10\degr$ of the centre of the average primary beam. A scaling factor was calculated from the ratio of the expected flux density at 154~MHz and the measured peak flux density in the MWA mosaic for each source. The mean of these scaling factors was 1.325 and the standard deviation was 19\% of the mean. The scaling factor was then applied to the MWA mosaic to produce the final, correctly scaled image. The likely cause for the initial lower-than-expected flux-density scale in the original MWA mosaic is the initial transfer of calibration solutions from a different antenna tile pointing, combined with the effects of self-calibration with an imperfect model. We take the 19\% standard deviation of the scaling factors as the uncertainty in the flux-density scale of the image.

\begin{figure*}
\centering 
\includegraphics[clip,trim=40 20 10 30,width=0.9\textwidth,angle=90]{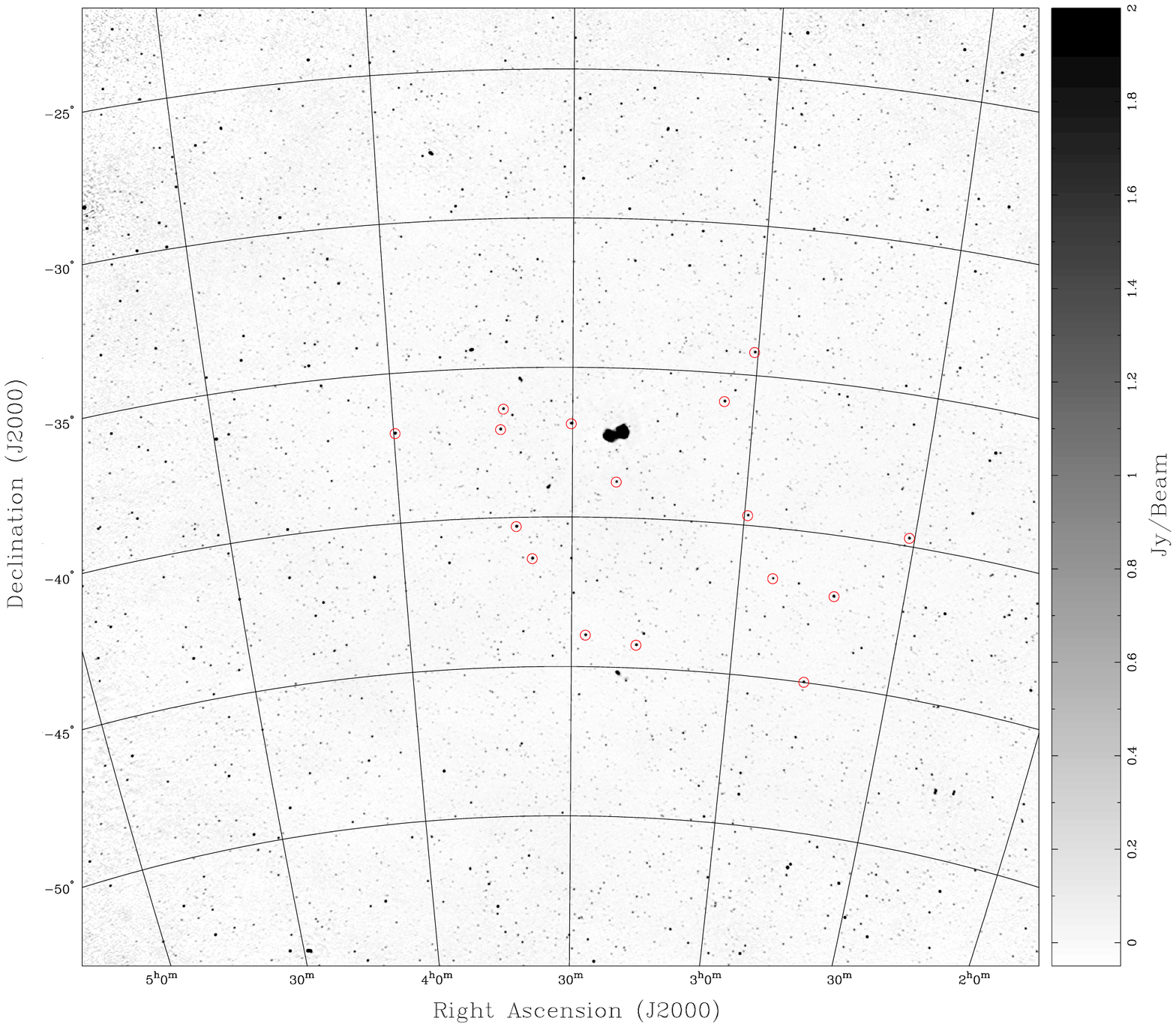}
\caption{Fornax~A and surrounding field at 154~MHz with the Murchison Widefield Array. The image is shown on a linear scale between $-0.05$ and $+2$ Jy/beam. It has an angular resolution of 185~arcsec and an rms noise of approximately 15~mJy/beam in `empty' regions of the image more than 1$\degr$ away from Fornax~A. The red circles mark the positions of the Culgoora sources  \citep{culgoora,culgoora1995} used to set the flux-density scale.}
\label{ForA_wide}
\end{figure*}

The final widefield image, with the Culgoora sources used to set the flux-density scale circled, is shown in Fig.~\ref{ForA_wide}. Fornax~A is clearly visible as the large double source near the centre of the image. The angular resolution is approximately 3~arcmin and the rms is approximately 15~mJy/beam in regions where there are no sources present above the 5-sigma level and more than 1$\degr$ away from Fornax~A. Close to Fornax~A the rms is slightly higher due to calibration and deconvolution errors and is approximately 25~mJy/beam. The peak brightness of the entire source is 19.7 Jy/beam, at a point in the west lobe at RA (J2000) 3\textsuperscript{h}21\textsuperscript{m}17\textsuperscript{s}, Dec (J2000) $-37\degree 9'10''$. The extent of the lobes is approximately 72~arcmin in the east-west direction, corresponding to a linear extent of 389~kpc.

\begin{figure*}
\centering 
\includegraphics[clip,trim=90 80 70 140,width=0.65\textwidth,angle=90]{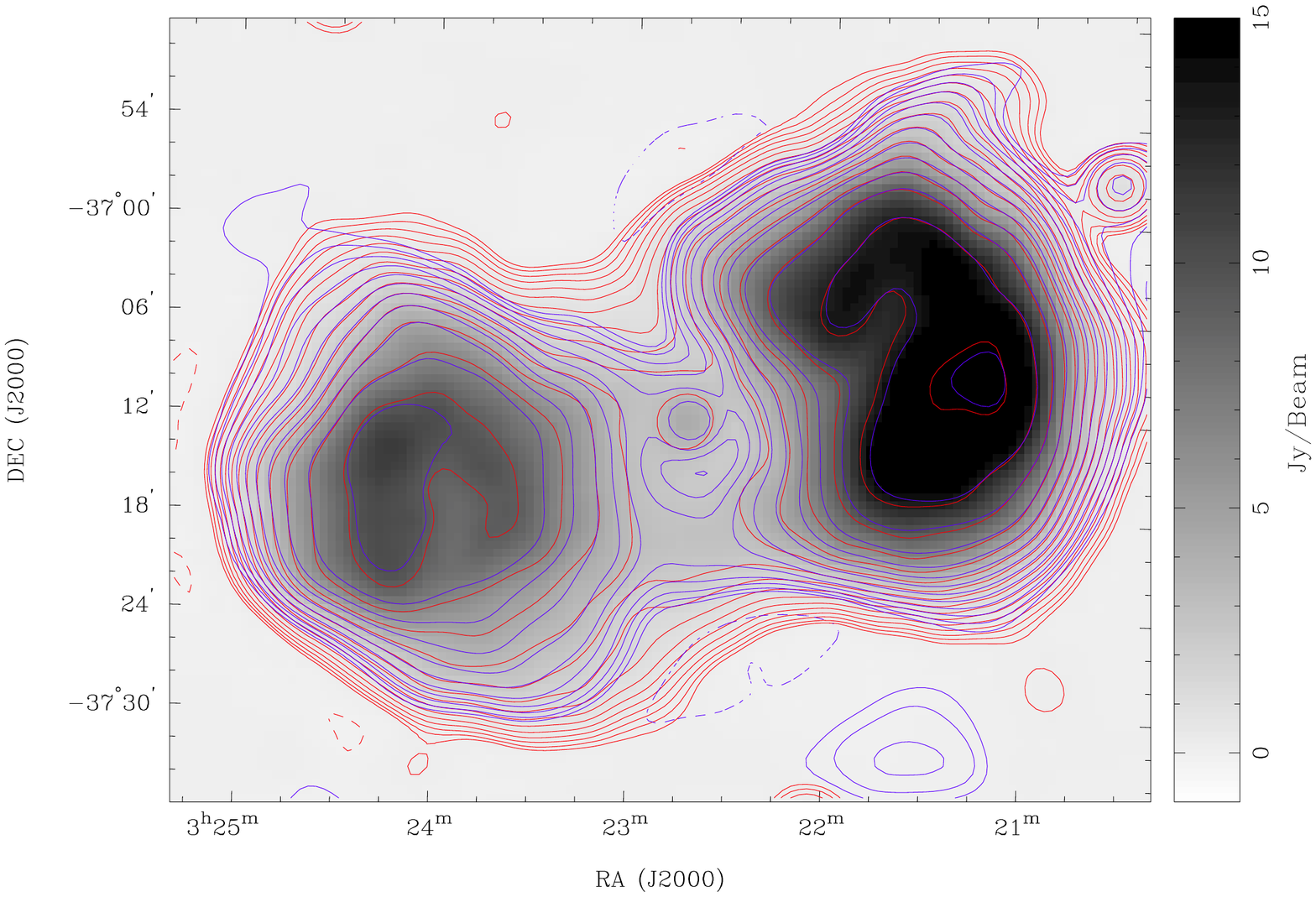}
\caption{Fornax~A at 154~MHz (gray scale and red contours) overlaid with the VLA 1.5~GHz \citep{fomalont} contours, after smoothing to the MWA resolution, in blue. Positive contours are solid and increment in a geometric progression of $\sqrt{2}$, starting at 0.1~Jy/beam for the 154~MHz image and 0.05~Jy/beam for the 1.5~GHz image. The broken red contour is at -0.1 Jy/beam and the broken blue contour is at -0.05~Jy/beam.}
\label{fig2}
\end{figure*}

The position of the peak pixel in the core is RA (J2000) 3\textsuperscript{h}22\textsuperscript{m}43\textsuperscript{s}, Dec (J2000) $-37\degree 12'2''$. This is consistent with the position of the host galaxy NGC~1316, given our pixel size of 28.8 arcsecs, which is located at RA (J2000) 3\textsuperscript{h}22\textsuperscript{m}41.718\textsuperscript{s}, Dec (J2000) $-37\degree 12'29.62''$ \citep{hubble_ForA}. The peak brightness of the core is 4.24~Jy/beam and the Gaussian restoring beam is $3.08\times3.08$~arcmin. Fig.~\ref{fig2} shows a zoomed in region of the MWA image at 154~MHz (gray scale and red contours) overlaid with the VLA 1.5~GHz image of \citet{fomalont}, which has been smoothed to the MWA resolution of approximately 3~arcmin. The image clearly shows the structure of the bright lobes, the fainter central core and a `bridge' of fainter diffuse emission present between the lobes to the north and south of the core. 

To measure flux densities for Fornax~A, we first blanked out all pixels with a value less than $0.20$~Jy/beam, then selected regions of the image using {\sc kvis} \citep{kvis}. We measured the flux density of the east lobe, defined as the region of the source east of RA (J2000) 3\textsuperscript{h}23\textsuperscript{m}0\textsuperscript{s} to be $260\pm50$~Jy. We defined the west lobe as the region to the west of RA (J2000) 3\textsuperscript{h}23\textsuperscript{m}0\textsuperscript{s}. This region also includes the compact core. Since the compact core remains unresolved at the MWA image resolution, we estimate its contribution to the flux-density of Fornax A as the peak brightness of the core, minus the brightness of the diffuse emission upon which it is superimposed. We estimate the brightness of the diffuse emission to be 2.27 Jy/beam, based on the minimum pixel value in the region just to the south-west of the core, which appears to be devoid of lobe emission. The core flux density is therefore $2.0\pm0.2$~Jy and the flux density of the west lobe is $490\pm90$~Jy. The total source flux density at 154~MHz is $750\pm140$~Jy. The errors are dominated by the 19\% flux-density scale uncertainty.

\subsection{Microwave-band data reduction}

We obtained flux-density measurements of Fornax~A at 30, 44, 70, 100 and 143~GHz by analyzing the most recently released {\it Planck} images \citep{planck_results_2013}, and at 23, 33, 41 and 61~GHz by analyzing the {\it WMAP} 9-year data release \citep{WMAP9}. We remove the contribution of the CMB by subtracting the {\it Planck} image at 217~GHz from the data. For the {\it WMAP} bands, the {\it Planck} 217~GHz image is degraded to the WMAP resolution (nside=512 in the Healpix \citep{healpix} pixelization scheme) before subtraction. For the {\it Planck} low-frequency bands (30, 44 and 70~GHz) the 217~GHz image is degraded to nside=1024 resolution before subtraction, and for the {\it Planck} high-frequency bands (100 and 143~GHz), the 217~GHz image is subtracted directly since these images are all at the same resolution.

We then make measurements at each frequency using aperture photometry, with an aperture size of 1 beam full-width-half-maximum (FWHM). The FWHM values were taken from \citet{planck_PCCS_2013} for {\it Planck} and from \citet{chen2013} for {\it WMAP}. Fornax~A transitions from being a single, unresolved source at the lower microwave frequencies, to being resolved into two distinct lobes at the higher microwave frequencies. Hence, for the measurements at 23, 30, 33, 41 and 44~GHz, we centre a single aperture on the coordinates of Fornax~A given in the {\it WMAP} 9-year catalogue \citep{WMAP9}, and for the measurements at 61, 70, 100, and 143~GHz, we centre an aperture on each of the lobes, with positions taken from \citet{planck_PCCS_2013}. An estimate of the background is subtracted from this integration and a correction factor is applied to take into account that the integration is truncated at 1 FWHM.

The flux density of foreground emission measured in the {\it Planck} images depends on the spectral index of the source, due to the way the data are calibrated and the frequency-dependent beam size. The {\it Planck} collaboration provides `colour corrections' to correct for this effect. Colour corrections for the {\it Planck} data were made by interpolating from the tables provided in \citet{PlanckLFI} for the low-frequency bands (30, 44 and 70~GHz) and in \citet{PlanckHFI} for the high-frequency bands (100 and 143~GHz). The spectral indices used for the colour corrections were determined from the images themselves by calculating the spectral indices between the bands. For the corrections at 30 and 44~GHz, we calculated the spectral index between these two bands and for the correction at 70~GHz, the spectral index between 44 and 70~GHz was used. These corrections were very small and within the photometric errors. No colour corrections were performed for the 100 and 143~GHz data, as the spectral indices computed were outside the range of the published colour correction table, however, \citet{HFImaps} explicitly state that the beam colour corrections at 100, 143 and 217~GHz are less than 0.3\%, which is negligible for our analysis.

The {\it WMAP} collaboration employ a different approach to colour corrections than {\it Planck}, where they compute an `effective frequency' for flux-density measurements of foregrounds, rather than a multiplicative factor for a particular frequency. We computed the effective frequencies for our {\it WMAP} measurements using the online calculator\footnote{http://lambda.gsfc.nasa.gov/product/map/dr5/effective\_freq.cfm}, which implements the equations for calculating the effective frequencies, as presented in \citet{jarosik}. The spectral indices used were those computed from the {\it WMAP} images, between adjacent frequency bands.

The conversions from T$_{\rm cmb}$ to Jy were made for the {\it Planck} measurements using the 2D Gaussian beam FWHMs and centre frequencies from Table 1 of \citet{planck_PCCS_2013} and, for the {\it WMAP} measurements, the computed effective frequencies and the beam FWHMs presented in  \citet{chen2013}. The flux-density values and effective frequencies of the {\it Planck} and {\it WMAP} planck data are listed in Table \ref{flux_table_all} and used in the SED modelling in Section 4.2.

\subsection{\emph{Fermi}-LAT data reduction}

We selected five years of data (MET 239557417 - MET 401341317) observed by {\it Fermi}-LAT for regions around Fornax~A and used the standard LAT analysis software (v9r32p5)\footnote{http://fermi.gsfc.nasa.gov/ssc}. The photons above $100~\rm MeV$ were selected for the analysis. The region-of-interest (ROI) was selected to be a $20^ \circ \times 20^ \circ$ square centred on the position of Fornax~A. To reduce the effect of the Earth's albedo background, time intervals when the Earth was appreciably in the field-of-view (FoV)\footnote{That is when the centre of the FoV is more than $52\degr$ from zenith, as well as time intervals when parts of the ROI are observed at zenith angles $>100\degr$.} were also excluded from the analysis. The spectral analysis was performed based on the P7REP version of post-launch instrument response functions. Both the front and back-converted photons were selected. The standard likelihood-analysis method, incorporated in the routine {\sc gtlike}, was adopted. 

\begin{figure*}
\centering 
\subfloat
{
\includegraphics[clip,trim=70 240 100 100,width=0.41\textwidth,angle=90]{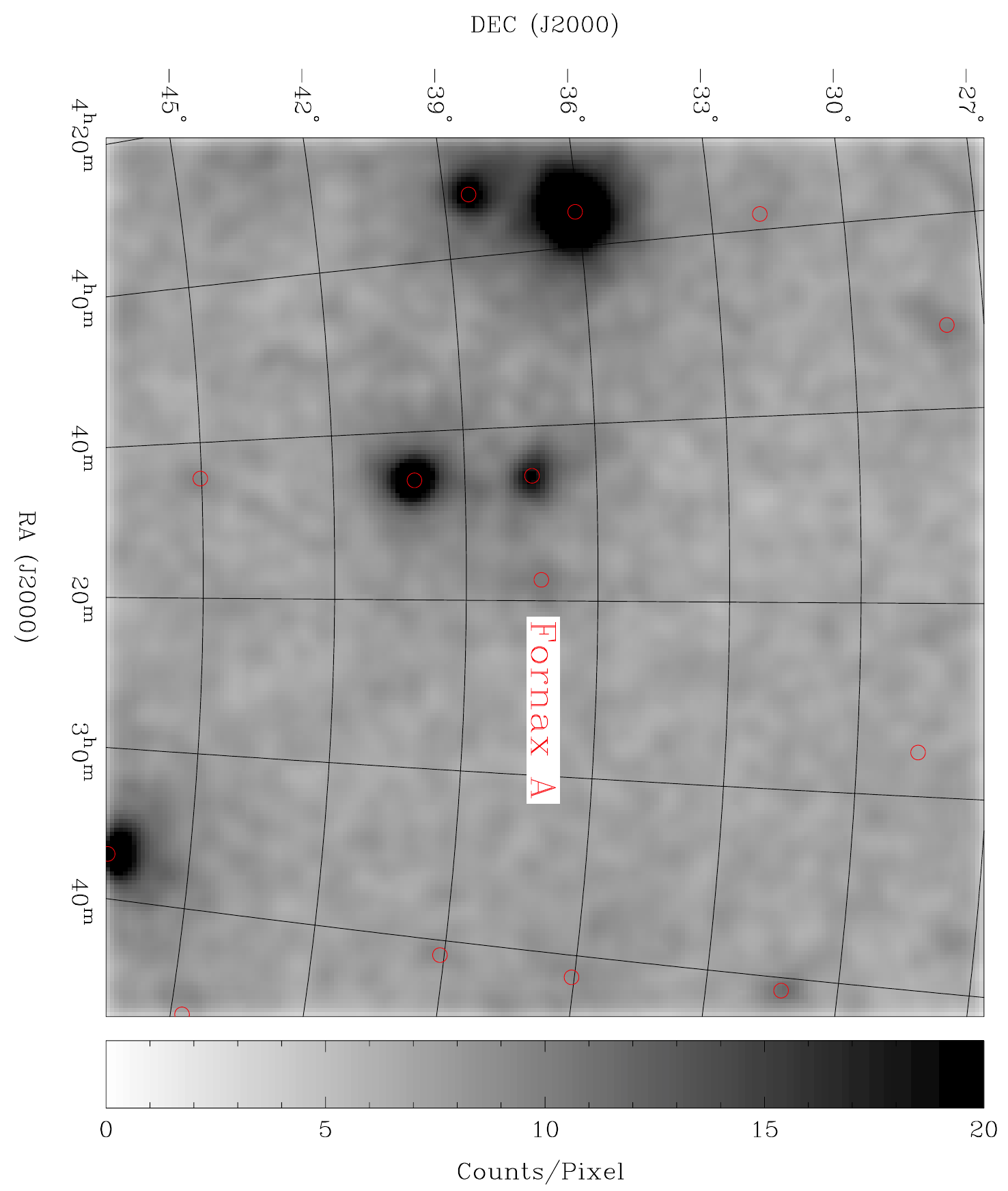}
\label{counts}
}
\subfloat
{
\includegraphics[clip,trim=130 280 40 60,width=0.41\textwidth,angle=90]{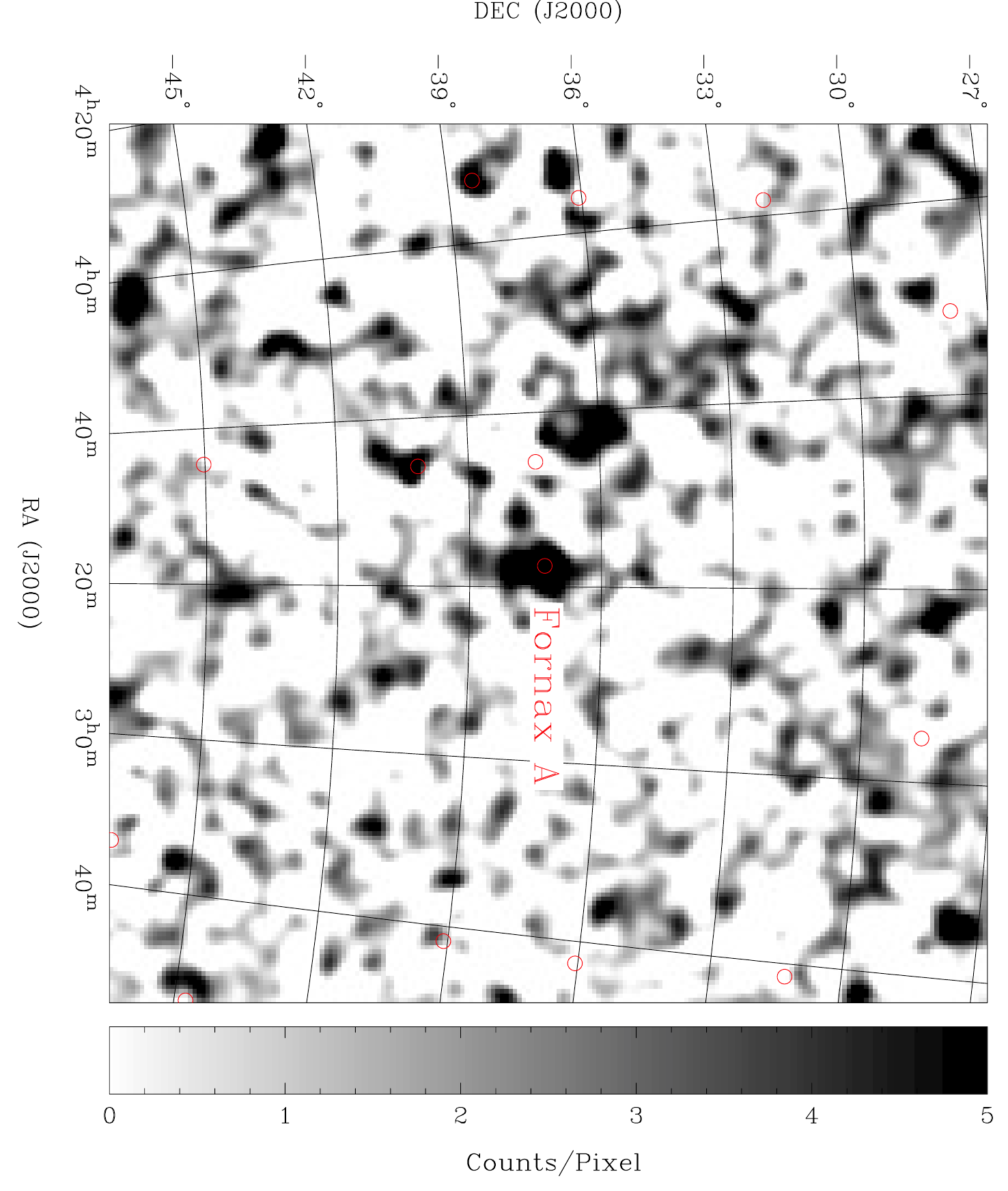}
\label{residuals}
}
\caption{$\gamma$-ray counts image (left panel) and residual image (right panel) above $100~\rm MeV$ in the ROI.  The sources in the 2FGL catalogue are marked as red circles.}
\label{fermi_maps}
\end{figure*}

The Galactic and isotropic diffuse models provided by the {\it Fermi} collaboration\footnote{Files: gll\_iem\_v05.fit and iso\_source\_v05.txt available at \\ http://fermi.gsfc.nasa.gov/ssc/data/access/\\lat/BackgroundModels.html} were used in the analysis. 2FGL sources \citep{2fgl} were also included and the parameters for point sources within $5\degr$ of Fornax~A were allowed to vary. The counts image and residual image above $100~\rm MeV$ are shown in Fig.~\ref{fermi_maps}. To study the spatial extension of the source, we also introduced disk templates and varied the radius of the disk, but found no improvement in the fitting. Thus, we treat Fornax~A as a point source in the spectral analysis.  It should be noted that Fornax~A is already included in the 2FGL catalogue \citep{2fgl} as a point source (2FGL J0322.4-3717). 

The best-fit position of Fornax~A was found to be RA (J2000) 3\textsuperscript{h}22\textsuperscript{m}34\textsuperscript{s}, Dec (J2000) -37\degree 21$'$11$''$, with an error radius of 0.13$\degree$, which is consistent with the position given in 2FGL catalogue of RA (J2000) 3\textsuperscript{h}22\textsuperscript{m}24\textsuperscript{s}, Dec (J2000) -37\degree 17$'$31$''$. We find that Fornax~A has a flux of $6.7\times 10^{-9}~\rm ph\ cm^{-2}\ s^{-1}$ and a photon index of $2.2 \pm 0.1$ above $100~\rm MeV$ (photon index, $\Gamma$, defined as ${\rm d}N/{\rm d}E\propto E^{-\Gamma}$, where $N$ is the number of photons and $E$ is energy). The test statistic value of the source was 64 in the same energy range, corresponding to a significance of about $8\sigma$. These values are consistent with the values for Fornax~A reported by \citet{2fgl} of $0.5\times10^{-9}~\rm ph\ cm^{-2}\ s^{-1}$ above 1000~MeV.

\begin{figure}
\centering 
\includegraphics[clip,trim=50 200 50 100,width=0.55\textwidth,angle=90]{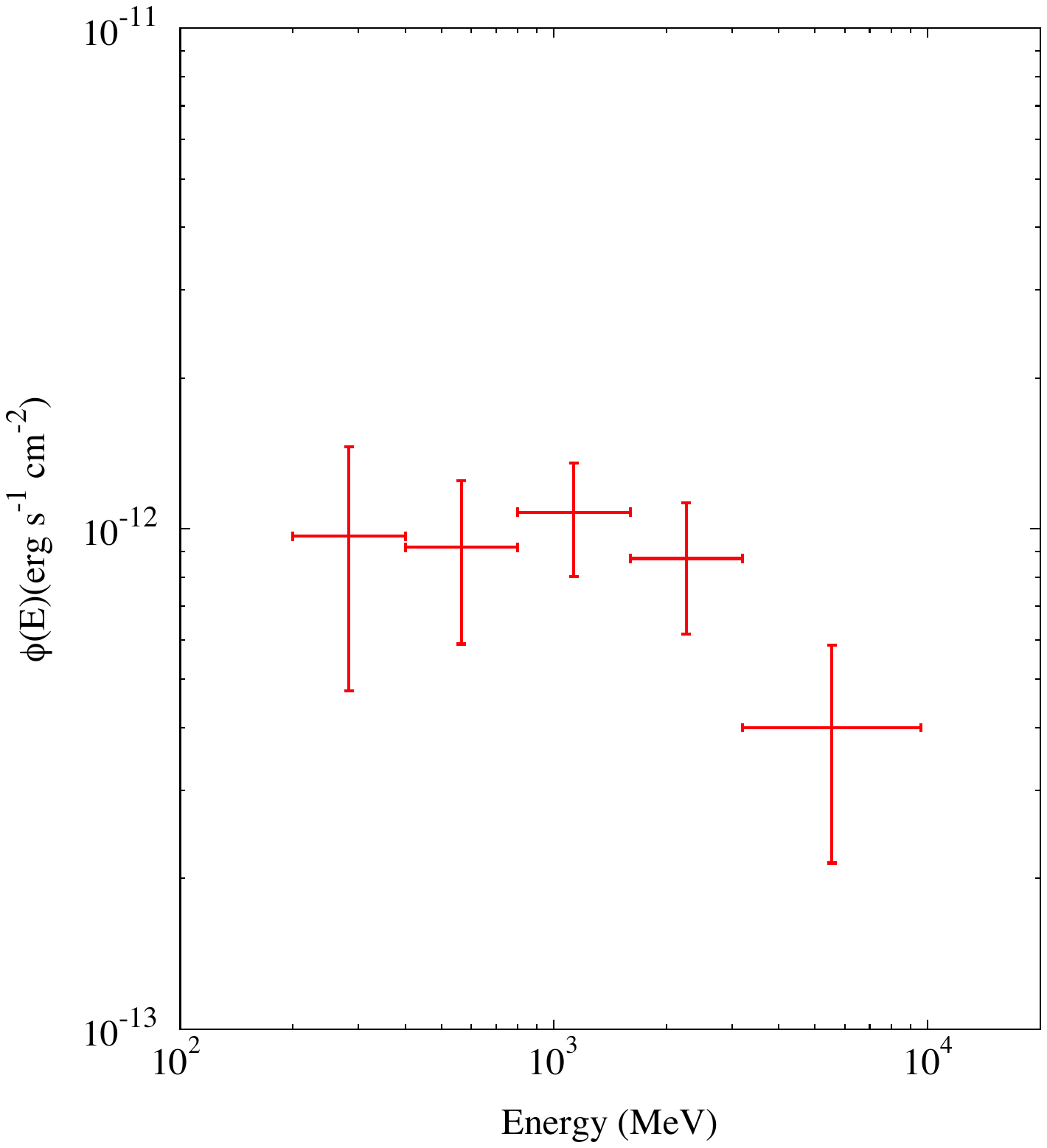}
\caption{The SED of Fornax~A above $100~\rm MeV$ derived from {\it Fermi}-LAT observations.}
\label{gamma_SED}
\end{figure}

To obtain the SED of Fornax~A above $100~\rm MeV$, we divided the energy range into logarithmically spaced bands and applied {\sc gtlike} in each band. Only the energy bins for which a signal was detected with a significance of at least $2 \sigma$ were considered. The derived high-energy SED is shown in Fig.~\ref{gamma_SED}. The $\gamma$-ray flux is approximately constant from 200 to 3000~MeV, but drops significantly in the highest energy bin.

\subsection{Previously published data}

We have conducted a thorough search of the literature and compiled a comprehensive list of flux-density measurements  of Fornax~A across a wide frequency range. These are summarised in Table~\ref{flux_table_all}. The listed flux-density measurements are for the whole source, since there are only a few published flux-density measurements of the individual lobes. Where the measurements are resolved, we report the sum of the two lobes and where the core is also measured, we include this in the sum, as it is included in all of the unresolved measurements. The inclusion of the core is a source of error, however, it is not considered to affect the results of our modelling significantly. At 154~MHz the core represents less than 2\% of the total flux density and at 1415~MHz the contribution of the core is less than 0.2\%. We show in Section 4.1.1 that the core has a steeper spectrum than the lobes so its contribution above 1415~MHz is negligible. In Table~\ref{flux_table_all} we list both the frequency, $\nu$, in MHz, and the equivalent energy, $h\nu$, in MeV. We also show the flux density, $S_{\nu}$, in Jy and convert this to a flux, $\phi$, in ($\times10^{-12}$) erg cm$^{-2}$ s$^{-1}$. For all measurements below 1~MeV the flux is calculated as $\phi=\nu S_{\nu}$, and for the $\gamma$-ray data we take into account the large logarithmically-spaced energy bin sizes ($E$), by using  $\phi=(h\nu)^2 {\rm d}N/{\rm d}E$, where $N$ is the photon count in cm$^{-2}$ s$^{-1}$.  Notes on the published data points are given in this section.

\begin{table*}
\caption{Fornax~A SED data, including the frequency, $\nu$, in MHz, the equivalent energy, $h\nu$, in MeV, the flux density of the whole source, $S_{\nu,{\rm total}}$ and where available the flux densities of the west and east lobes individually ($S_{\nu,{\rm west}}$, $S_{\nu,{\rm east}}$) in Jy and the corresponding flux, $\phi$, in  ($\times10^{-12}$) erg cm$^{-2}$ s$^{-1}$.  In cases where the flux densities of the individual lobes do not add up to the total flux density, the remaining flux density resides in the core.}
\label{flux_table_all}
\begin{tabular}{@{}lcccccc} 
 \hline
 \shortstack{$\nu$\\(MHz)} &  \shortstack{$h\nu$\\(MeV)} & \shortstack{$S_{\nu,{\rm total}}$ ($S_{\nu,{\rm west}}$, $S_{\nu,{\rm east}}$)\\(Jy)} & \shortstack{$\phi$ \\ ($\times10^{-12}$ erg  s$^{-1}$ cm$^{-2}$)} & \shortstack{Uncertainty} & \shortstack{Reference } \\
 \hline
4.7* & 1.94$\times10^{-14}$  &  13500 & 0.634 & 20\% & \citet{ForA_5MHz}  \\
18.3  & 7.57$\times10^{-14}$ &  3500 & 0.640 & 20\% & \citet{ForA_18MHz} \\
19.7 & 8.15$\times10^{-14}$ &  4300 &  0.847 & 20\% & \citet{ForA_20MHz} \\
29.9 & 1.24$\times10^{-13}$  &  2120 & 0.634 & 10\% & \citet{ForA_29MHz} \\
85.7 & 3.54$\times10^{-13}$ & 950 &  0.814 & 20\% & \citet{ForA_85MHz} \\
100*  & 4.14$\times10^{-13}$ &  200,240 & 0.240 & 20\% & \citet{ForA_100MHz,ForA_100MHz_2} \\
154  & 6.37$\times10^{-13}$ &  750 (478,260) & 1.16 &19\% & This work  \\
189* & 7.82$\times10^{-13}$ & 519 &  0.981 & 5\% & \citet{bernardi}\\
400*  & 1.65$\times10^{-12}$ & 140 &  0.560 & 10\% & \citet{ForA_400MHz} \\
408  & 1.69$\times10^{-12}$ & 259 &  1.06 & 10\% & \citet{ForA_408MHz,cameron} \\
600  & 2.48$\times10^{-12}$ & 310 &  1.86 & 25\% & \citet{ForA_600MHz} \\
843  & 3.49$\times10^{-12}$ & 169 &  1.42 & 9\% & \citet{ForA_843MHz} \\
1415  & 5.85$\times10^{-12}$ & 125  & 1.77 &8\% & \citet{ekers}  \\
1510 & 6.24$\times10^{-12}$   & 117 (77.2, 39.2) &  1.77 & 10\% & \citet{fomalont} \\
2700  & 1.11$\times10^{-11}$  & 98 (67, 31) &  2.65 & 10\% & \citet{shimmins} \\
5000   & 2.07$\times10^{-11}$ &  54.7 (31.5, 23.2) &  2.74 & 15\% & \citet{5GHz} \\
22520 & 9.51$\times10^{-11}$ & 14.1 &  3.18 & 5\% & This work \\
28400 & 9.51$\times10^{-11}$ & 10.9 &  3.10 & 7\% & This work \\
32750 & 1.36$\times10^{-10}$ & 9.4 &  3.08 & 6\% &This work \\
40430  & 1.70$\times10^{-10}$ & 6.6 &   2.67 & 9\% &This work \\
44100  & 1.70$\times10^{-10}$ & 6.1 &   2.70 & 20\% &This work \\
60350  &  2.52$\times10^{-10}$ & 4.7 (2.6, 2.1) &  2.84 & 15\% & This work \\
70400  & 2.89$\times10^{-10}$  & 3.0 (1.6, 1.4) & 2.11 & 17\% & This work \\
100000  & 4.14$\times10^{-10}$ & 1.2 (0.60, 0.60) &  1.20 & 25\% & This work\\
143000  & 5.91$\times10^{-10}$ & 0.29 (0.19, 0.10) &  0.415 & 51\% & This work\\
2.4$\times10^{+11}$ & 1.00$\times10^{-03}$ & 2.06$\times10^{-7}$ &  0.498  & 20\% & \citet{tashiro,isobe} \\
6.8$\times10^{+16}$  & 2.83$\times10^{+02}$ & 1.4$\times10^{-12}$ &  0.968 & 51\% & This work \\
1.4$\times10^{+17}$ & 5.66$\times10^{+02}$ & 6.7$\times10^{-13}$ &  0.919 & 36\% &This work \\
2.7$\times10^{+17}$ & 1.13$\times10^{+03}$ &  3.9$\times10^{-13}$ & 1.08  & 26\% &This work\\
5.5$\times10^{+17}$ & 2.26$\times10^{+03}$ & 1.6$\times10^{-13}$ &  0.873 & 29\% & This work \\
1.3$\times10^{+18}$ & 5.54$\times10^{+03}$ & 3.0$\times10^{-14}$ &  0.401 & 46\% &This work \\
\hline
\multicolumn{6}{l}{*These data have not been included in the SED fitting in Section 4.2 for reasons detailed in the text.}
\end{tabular}\\
\end{table*}

\subsubsection{Low-frequency data: 4.7$-$843~MHz }

Many of the observations of Fornax~A below 1~GHz come from the very early days of radio astronomy, when the first radio interferometers were being built and used to discover discrete radio sources in the sky for the first time. Many of these observations had poor angular resolution by today's standards and large flux-scale uncertainties. However, they are sufficient to constrain SED models of Fornax~A over the large energy ranges that are being considered in this paper. The flux-density measurements of Fornax~A, their estimated uncertainties, and references for each measurement are given in Table~\ref{flux_table_all}. Not all references explicitly state the estimated uncertainties in their measurements, so we have used the estimate of a 20\% error in the overall flux-density scale in the \citet{ForA_20MHz} measurements as a guide when estimating errors in flux densities reported in this early era of radio astronomy. We consider that the flux-density measurements at 100~MHz (\citealt{ForA_100MHz,ForA_100MHz_2}) and the measurement at 400~MHz \citep{ForA_400MHz} are unreliable, due to probable systematic errors in the overall flux-scale calibration, and we do not include them in our SED fitting in Section 4.2. We also do not include the 189~MHz measurements of \citet{bernardi} in our SED fitting, since our observed frequency is very close to theirs and we consider our 128-tile measurements to be an improvement on the MWA~32T results.

\subsubsection{GHz-frequency data: 1.4$-$5~GHz }

We use the flux-density measurements of the lobes and core of Fornax~A from \citet{ekers} at 1415~MHz. The observations use data from the Fleurs Synthesis Telescope, combined with 1.4~GHz data from the Parkes radio telescope to fill in the central part of the uv plane.

We measure the flux densities of the lobes and the unresolved core of Fornax~A at 1.5~GHz using the 14-arcsec resolution image of \citet{fomalont}. Details of the observations used by \citet{fomalont} are not given in their paper, however, the reasonably good agreement between the flux densities we measure in the \citet{fomalont} image at 1510~MHz and those reported by \citet{ekers} indicates that, despite its high angular resolution, the 1.5~GHz image is not missing significant emission on the angular scales of the lobes and therefore our flux-density measurements are reliable. 

We obtained flux-density measurements for the west and east lobes of Fornax~A from the Parkes 2700~MHz survey \citep{shimmins}. 

We measured the flux density at 5~GHz for both of Fornax~A's lobes from the Parkes image of \citet{5GHz}. A software planimeter\footnote{http://www.csudh.edu/math/sraianu/} was used to compute the area enclosed by each contour and the flux density was calculated from this area and the brightness temperature indicated by the contour. A background contribution of 0.78~K across the source was subtracted from the integrated flux density. We measured flux densities of $31.5\pm4.7$~Jy and $23.2\pm3.4$~Jy for the west and east lobes respectively, giving a total source flux density of $54.7\pm8.2$~Jy, which is consistent with the value of 49~Jy reported by \citet{kuhr}, who do not report separate values for each of the lobes or provide an image.

\subsubsection{X-ray data} 

We use the X-ray flux densities reported by \citet{tashiro} and \citet{isobe} for the west and east lobes, respectively, at 1~keV. Here we assume that the reported flux densities represent the lobe emission with negligible contribution from thermal emission. We expect this to be a valid assumption since \citet{tashiro} model and remove the thermal emission, and while \citet{isobe} do not account for thermal emission, their data is very well fit by a power-law model at $>3$ keV,  where the thermal emission should be negligible, given their best-fit temperature of less than 1~keV.

\section{Spectral Energy Distribution Analysis}

\subsection{Spectral index between 154~MHz and 1.5~GHz}

Since we have images at 154~MHz and 1.5~GHz with reasonable angular resolution, we investigate the spectral index between these two frequencies in detail. Fig.~\ref{fig2} shows that the morphologies at both frequencies match closely for the lobes, however, there are clear differences in the shape of the contours in the bridge region between the lobes and to the north and south of the compact, unresolved core. The core itself is much more prominent at 154~MHz than at 1510~MHz. 

We expect from the previous IC modelling of \citet{tashiro} and \citet{isobe} that there is little curvature in the spectrum of the lobes between these two frequencies, and the relatively large fractional frequency coverage allows us to accurately measure the spectral index. The spatial distribution of the spectral index across the source is of interest as it provides clues about the evolutionary history and the physical properties of the radio galaxy. At the 3-arcmin angular resolution of Fig. \ref{fig2}, the close match between the MWA and VLA contours indicates that there is little spatial variation in the spectral index across the lobes. At this resolution, however, we are unable to resolve the filamentary structure of the lobes shown by \citet{fomalont}. We investigate the spatial variation of the spectral index of Fornax~A at the MWA resolution, using spectral tomography and spectral-index mapping, in Sections 4.1.1 and 4.1.2, respectively.

\subsubsection{Spectral tomography}

We use the technique of spectral tomography to investigate the spatial variation of the spectral index over the source. This technique is useful for identifying regions of different spectral indices in complex structures, such as the lobes of radio galaxies, where there may be distinct structures that overlap in our line-of-sight that are difficult to identify using traditional spectral-index maps (see e.g. \citealt{katz-stone,HerA,gaenslerwallace,mckinley}). Since we have already identified that the Fornax A core appears to be superimposed on a bridge of diffuse emission, we use spectral tomography here to attempt to disentangle the spectral indices of these structures, and to determine if there are any other regions of the source with overlapping spectral-index components.

\begin{figure*}
\centering 
\includegraphics[clip,trim=210 210 200 55,width=0.32\textwidth,angle=90]{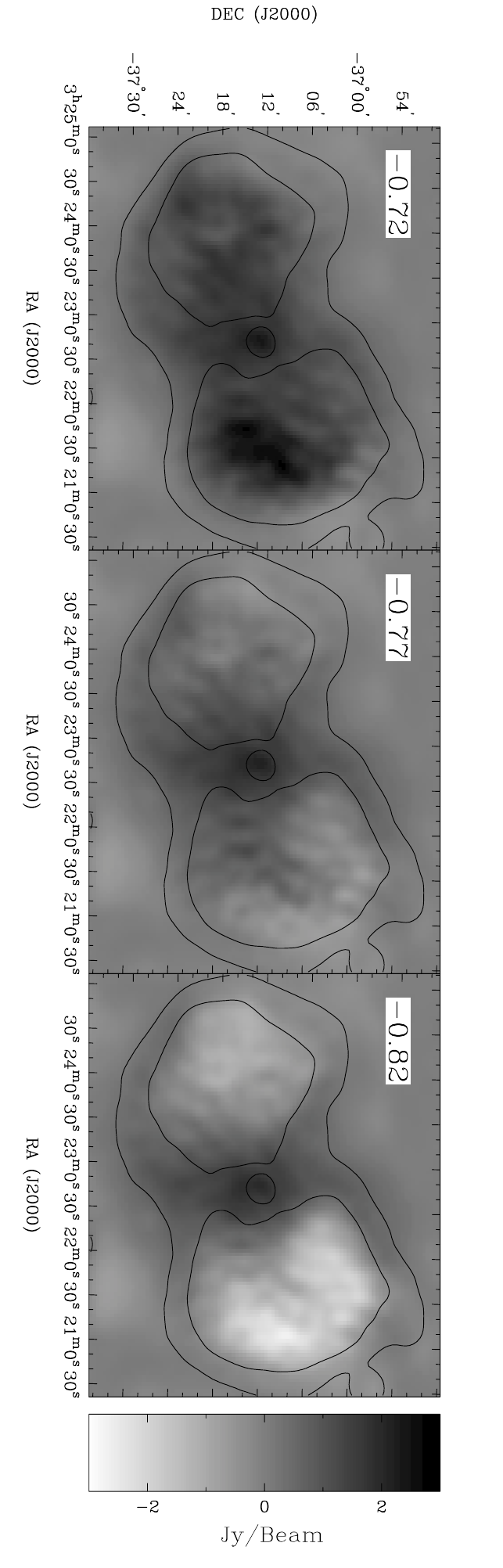}
\caption{Slices from the spectral tomography cube, with trial spectral indices indicated in the top left corner of each panel, showing that the spectral index of both lobes is between $-0.72$ and $-0.82$. The gray scale is linear and from -3 to 3 Jy/beam. The contours are from the MWA 154~MHz image at 0.2 and 3.0 Jy/beam.}
\label{tomography_lobes}
\end{figure*}

We constructed a tomography cube using our 154~MHz image and the 1.5~GHz VLA image \citep{fomalont}, following the procedure described by \citet{mckinley}. Fig.~\ref{tomography_lobes} shows three slices from the spectral tomography cube, with trial spectral indices, $\alpha_t$, as indicated in the top-left corner of each panel. Fig.~\ref{tomography_lobes} shows both lobes to be completely under-subtracted at $\alpha_t=-0.72$ and completely over-subtracted at  at $\alpha_t=-0.82$. We therefore estimate the spectral index of both lobes to be $\alpha_r=-0.77 \pm 0.05_{\rm{stat}}$, where the quoted error is the statistical error only, since the systematic errors are not relevant for comparing relative spectral indices across the source. We find no strong evidence for overlapping structures with different spectral indices in the lobes. The low-level ripple structure present in the lobes in Fig.~\ref{tomography_lobes} is thought to be due to minor deconvolution errors associated with using delta functions to represent diffuse structure. This could possibly be overcome by using alternative techniques such as multi-scale clean or maximum entropy method deconvolution, however we leave experimentation with these algorithms for future work, since the errors do not impact on the main scientific results of this paper. 

\begin{figure*}
\centering 
\includegraphics[clip,trim=210 200 200 48,width=0.32\textwidth,angle=90]{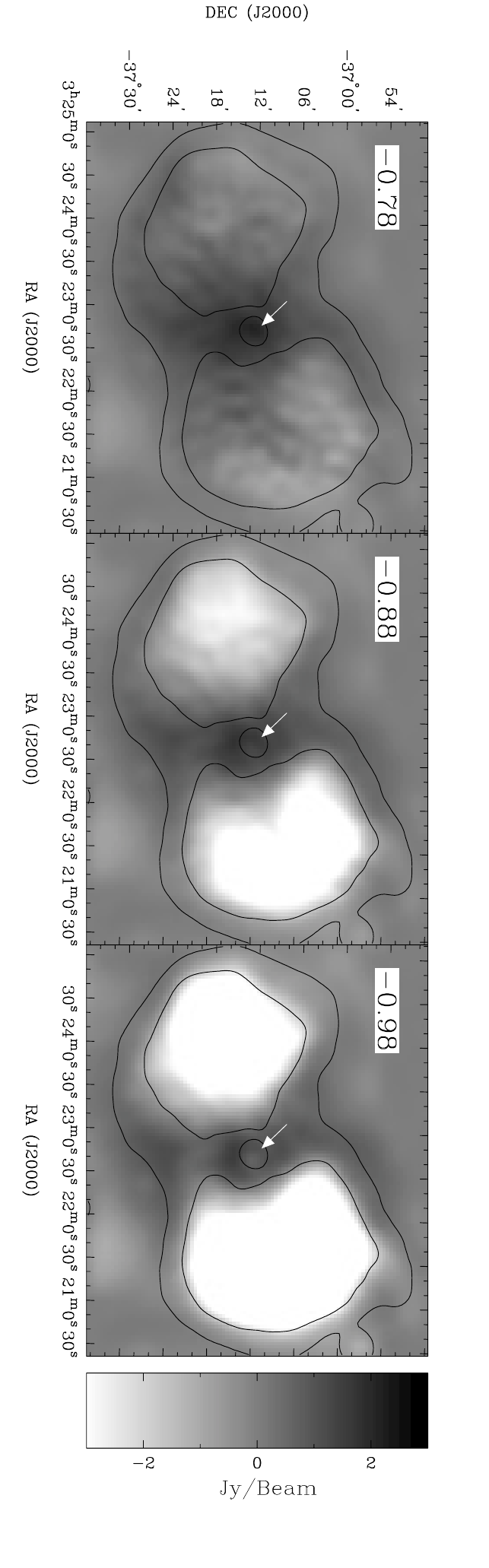}
\caption{As for figure 6, but with trial spectral indices between $-0.78$ and $-0.98$. The unresolved core of Fornax~A, indicated by the white arrow, has a flatter spectrum than the surrounding diffuse bridge emission.}
\label{tomography_core}
\end{figure*}

The core region, labelled with an arrow in the three panels of Fig.~\ref{tomography_core}, has a somewhat steeper spectrum than the lobes, which we estimate at $\alpha_{core} = -0.88 \pm 0.10_{\rm{stat}}$. The spectrum of the core is, however, significantly flatter than the surrounding diffuse bridge emission. This central region of Fornax A appears to be complex, with overlapping spectral-index contributions from the core, bridge and lobes. There does appear to be a region of the bridge free from overlapping structures, which is directly adjacent to the core in the south. We measure the spectrum of this `clean' region of the bridge in Section 4.1.2, using a traditional spectral-index map.

\subsubsection{Spectral-index mapping}

\begin{figure*}
\centering 
\includegraphics[clip,trim=110 240 50 30,width=0.75\textwidth,angle=90]{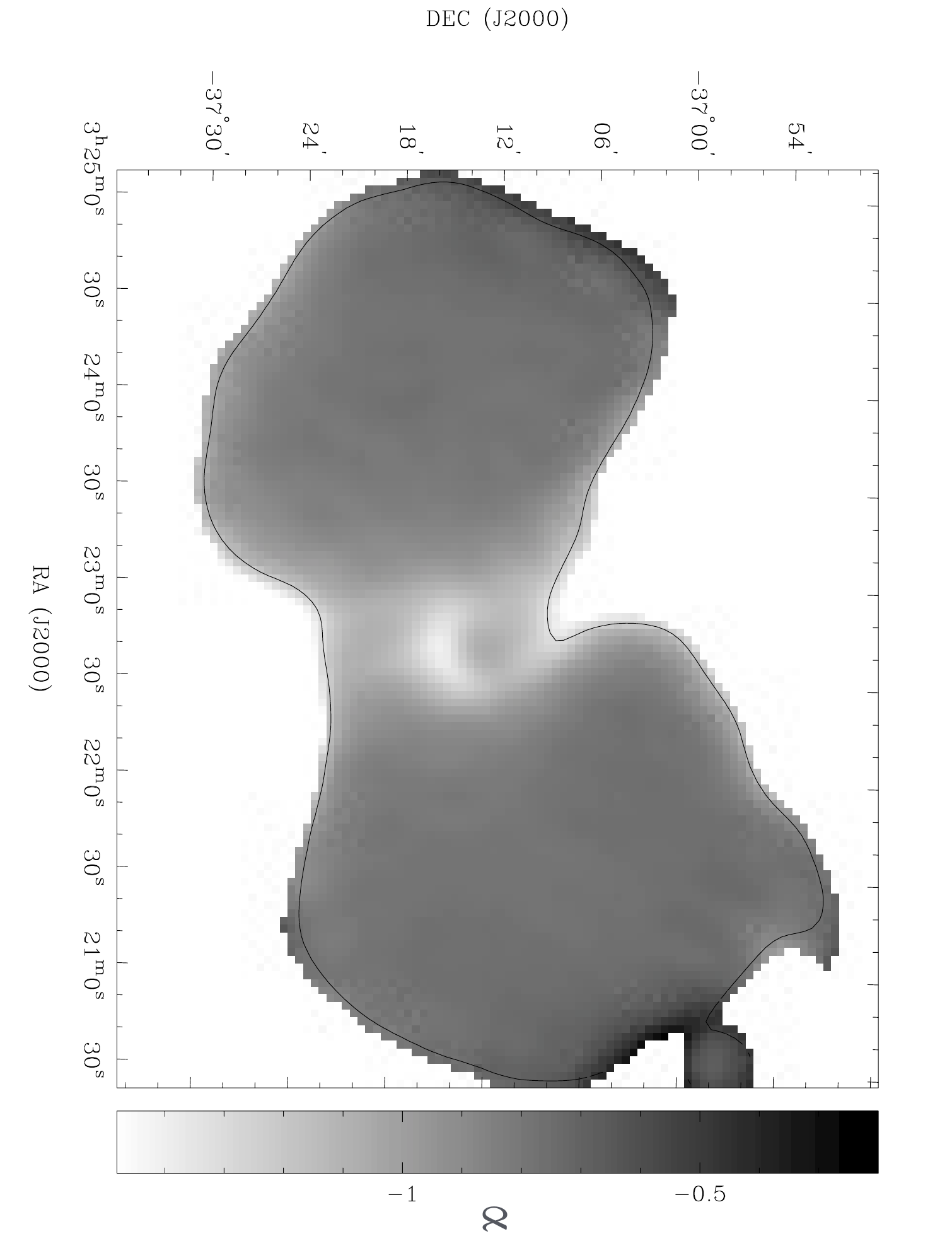}
\caption{Spectral index map of Fornax~A. The contour shown is the 0.1 Jy/beam level from the smoothed VLA image at 1510~MHz. White pixels outside of the contour are blanked and there are no blanked pixels inside the contour. The map shows that the spectral index varies only slightly across the lobes, from the average value of approximately $-0.77$, but that the bridge region has a significantly steeper spectrum; reaching a value of $-1.4$ just to the south of the core.}
\label{spec_index_map}
\end{figure*}

Fig. \ref{spec_index_map} shows the radio spectral index, $\alpha_r$, calculated between 154 and 1510~MHz, for each pixel in the 3-arcmin resolution images. The spectral index was only computed for pixels where the 154~MHz pixel was greater than 0.2~Jy/beam and the 1510~MHz pixel was greater than 0.05 Jy/beam. These spectral index values are all affected by the same systematic errors, which are dominated by the 19\% uncertainty in the flux-density scale of the MWA image. The spectral-index map shows that both lobes have an average spectral index of around $-0.77^{+0.09_{\rm sys}}_{-0.08_{\rm sys}}$, in agreement with the spectral tomography results. The bridge region clearly has a much steeper spectral index, reaching a value of $-1.4^{+0.09_{\rm sys}}_{-0.08_{\rm sys}}$ in the region just to the south of the core. The unresolved core, which has a flatter spectrum than the surrounding bridge, appears to be offset to the north of the centre of the bridge. 

\begin{figure*}
\centering 
\includegraphics[clip,trim=150 250 20 20,width=0.7\textwidth,angle=90]{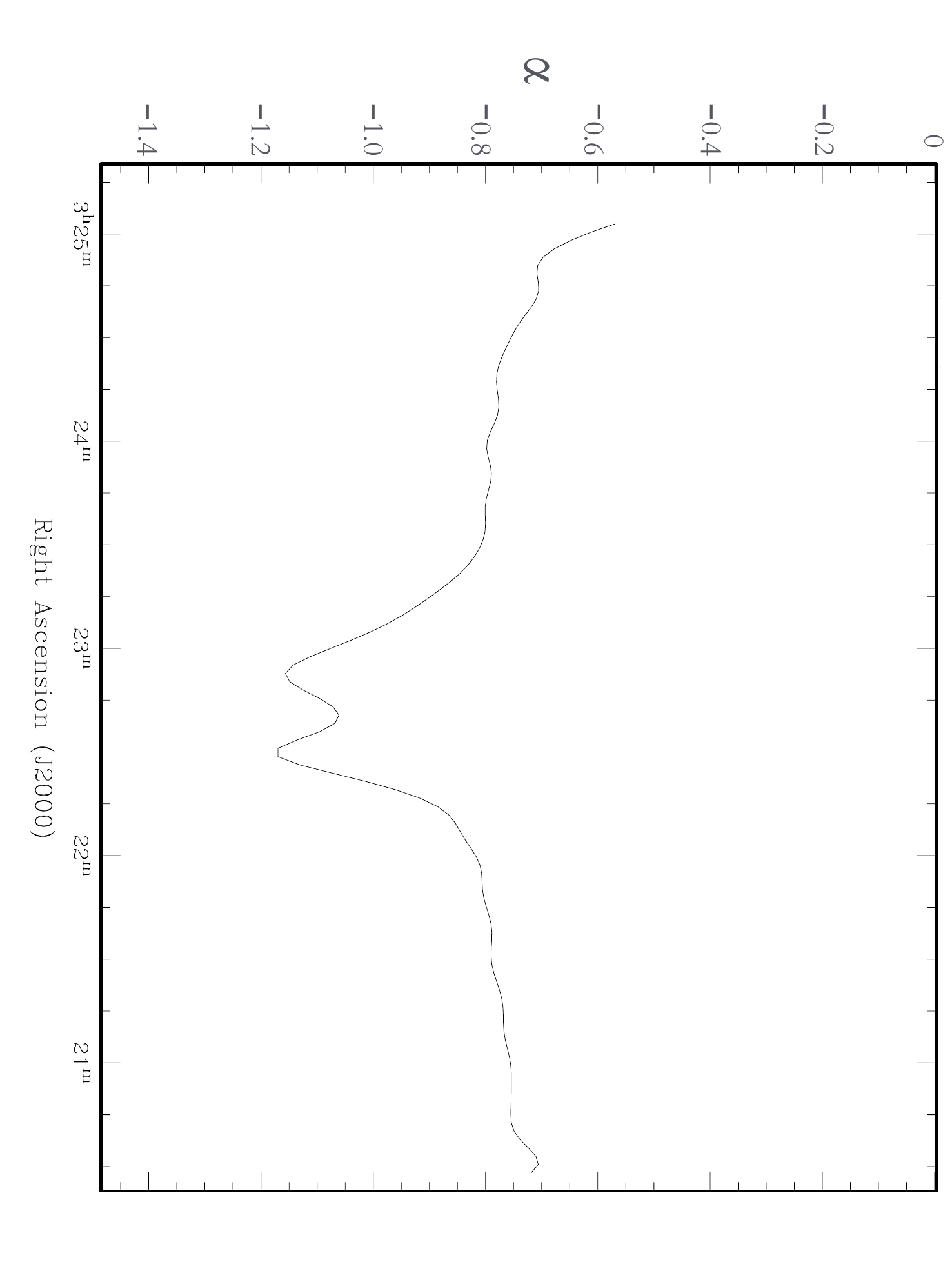}
\caption{Profile through the spectral index map of Fornax~A at DEC (J2000) $-40\degree 46\arcmin$.}
\label{spec_index_profile}
\end{figure*}

The spectral index of the source appears to steepen from the outer regions of the lobes in towards the core. This steepening of the spectrum towards the core is shown more clearly in Fig. \ref{spec_index_profile}, where we plot a profile of the spectral index through the source along a line at Dec (J2000) $-40\degree 46'$. The spectral index steepens gradually from around $-0.7$ at the edges of the lobes to $-0.8$ nearer to the core, before steepening rapidly across the bridge region. The spectrum of the core, appearing as a local maximum near the centre of the profile in Fig. \ref{spec_index_profile}, has a spectrum that is clearly steeper than the lobes, but flatter than the bridge. It must be noted, however, that the values of the spectral-index in the central regions of both Figs \ref{spec_index_map} and \ref{spec_index_profile} are a superposition of the steep bridge emission and the flatter, overlapping core and lobe emission.

\subsection{SED Fitting}

The {\it Fermi}-LAT observations have insufficient angular resolution to resolve the two Fornax~A lobes, however, we have shown in our analysis of the radio spectrum that the spectral behaviour of both lobes is very similar. As shown in Section 4.1, the spectral index between 154 and 1510~MHz, of both the west and east lobes, is $\alpha_r=-0.77\pm 0.05_{\rm{stat}}$. As such, we proceed with the IC modelling using the flux densities of the whole source listed in Table~\ref{flux_table_all}.

As noted in Section 3.5.1, we have excluded the points at 100 and 400~MHz in the SED fitting, due to probable systematic errors in these measurements. We have also excluded the point at 4.7~MHz as we found that the data below 5~GHz is fit well by a single power-law spectrum, but the 4.7~MHz point sits well above the expected flux density from this power-law model, which is difficult to physically explain and is more likely to be due to incorrect calibration of the overall flux-density scale, or other problems that are inherent in such very low-frequency observations.

To investigate the origin of the high energy emission in the Fornax~A lobes, we compare three models (I, II, III) for the broadband SED. In all three models we include both the CMB and the EBL (using the model of \citealt{EBL1}) as the seed photon sources for the IC scattering component. The best-fit model parameters are calculated by numerical minimization with {\sc pyminuit}. Each data point represents an independent measurement at that particular frequency/energy, allowing us to perform a $\chi^2$ analysis. The derived model parameters and the $\chi^2$ per degree of freedom for each fit are shown in Table~\ref{model_parameters}.

\begin{figure*}
\centering 
\includegraphics[clip,trim=0 0 0 0,width=1.0\textwidth,angle=0]{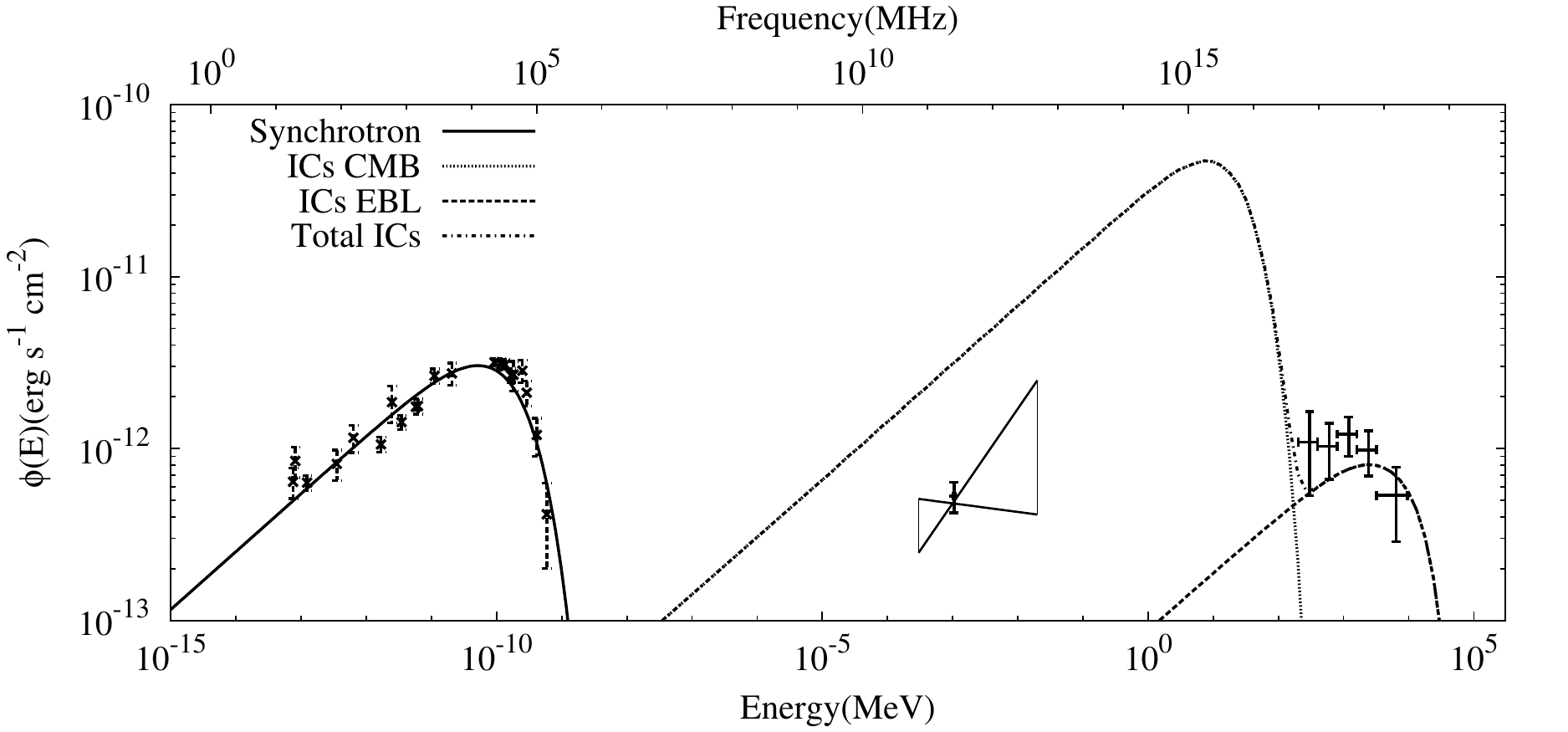}
\caption{The broadband SED of Fornax~A for Model~I, where both the X-rays and $\gamma$-rays are produced by IC scattering by the synchrotron-emitting electrons. The IC scattering contribution from the EBL and the CMB photon fields are both shown in the figure. The model parameters are described in Table~\ref{model_parameters}.}
\label{ForA_lep_01}
\end{figure*}

Model~I is a purely leptonic scenario, where we assume that both the X-rays and $\gamma$-rays are due to IC scattering of the seed photons by the synchrotron-emitting population of electrons, which have a distribution described by $N_e(E)=K (E/1{\rm GeV)}^{p} e^{-(E/E_{ce})^4}$, where $K$ is a normalisation constant, $E$ is energy, $p$ is the electron energy index and $E_{ce}$ is the high cutoff energy. We note that, in order to fit the unusually sharp cutoff in the synchrotron spectrum at the highest {\it Planck} frequencies, it is necessary to model the electron energy distribution with a super-exponential cutoff, $e^{-(E/E_{ce})^4}$. We are not aware of any physical process that would cause the energy distribution to cutoff so sharply. It is possible that this sharp cutoff at the higher frequencies is due to a non-zero flux density of Fornax~A in the 217~GHz image subtracted from the microwave data as a CMB template. At 217~GHz the total flux density of Fornax~A was measured to be $0.17 \pm 0.10$ Jy. The measurement is dominated by noise and the lobes are hardly visible by eye. This non-zero flux density would only significantly affect the measurements at 100 and 143~GHz, producing the sharp cutoff, and since the spectrum has clearly stopped rising by around 10~GHz, the conclusions of this work are not affected. The modelling results are shown in Fig. \ref{ForA_lep_01}. In this scenario, the IC scattering of the EBL accounts for the $\gamma$-ray emission well, but the IC scattering of the CMB results in an X-ray flux that is an order of magnitude more than the observed value at 1~keV.

\begin{figure*}
\centering 
\includegraphics[clip,trim=0 0 0 0,width=1.0\textwidth,angle=0]{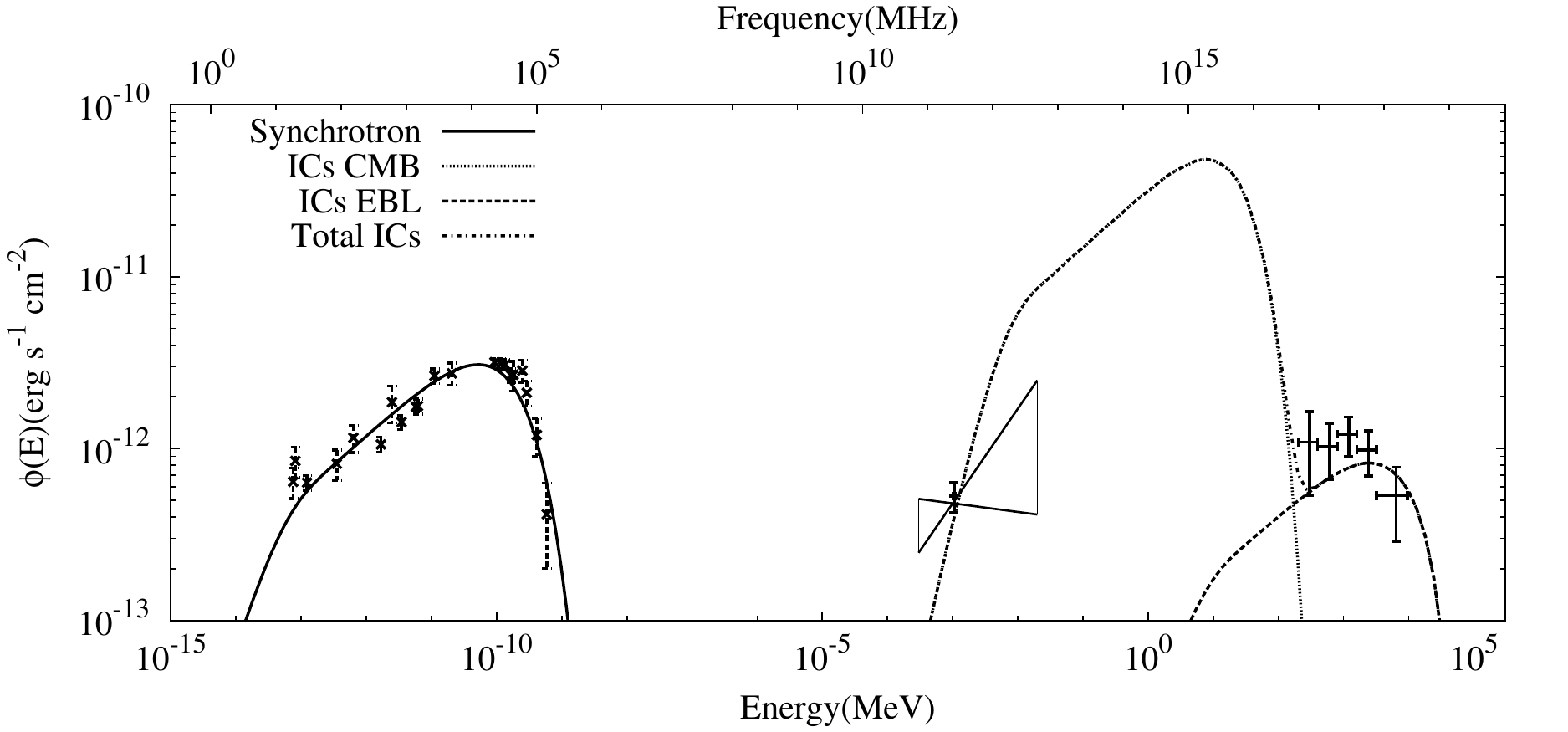}
\caption{The broadband SED of Fornax~A for Model~II, where both the X-rays and $\gamma$-rays are produced by IC scattering by the synchrotron-emitting electrons, which are described by an electron energy distribution with a low-energy cut-off. The model parameters are described in Table~\ref{model_parameters}.}
\label{ForA_lep_02}
\end{figure*}

Model~II is the same as Model~I, but we introduce a low-energy cutoff in the electron energy distribution, such that $N_e(E)=0$ for $E<0.9$ GeV. The modelling results are shown in Fig. \ref{ForA_lep_02}. In this case the model fits the high energy data points, but the X-ray spectral index at 1~keV is approximately zero, which is much harder than the observed values of $-0.62^{+0.15}_{-0.24}$ \citep{isobe} and $-0.81\pm0.22$ \citep{tashiro} for the east and west lobes, respectively.

\begin{figure*}
\centering 
\includegraphics[clip,trim=0 0 0 0,width=1.0\textwidth,angle=0]{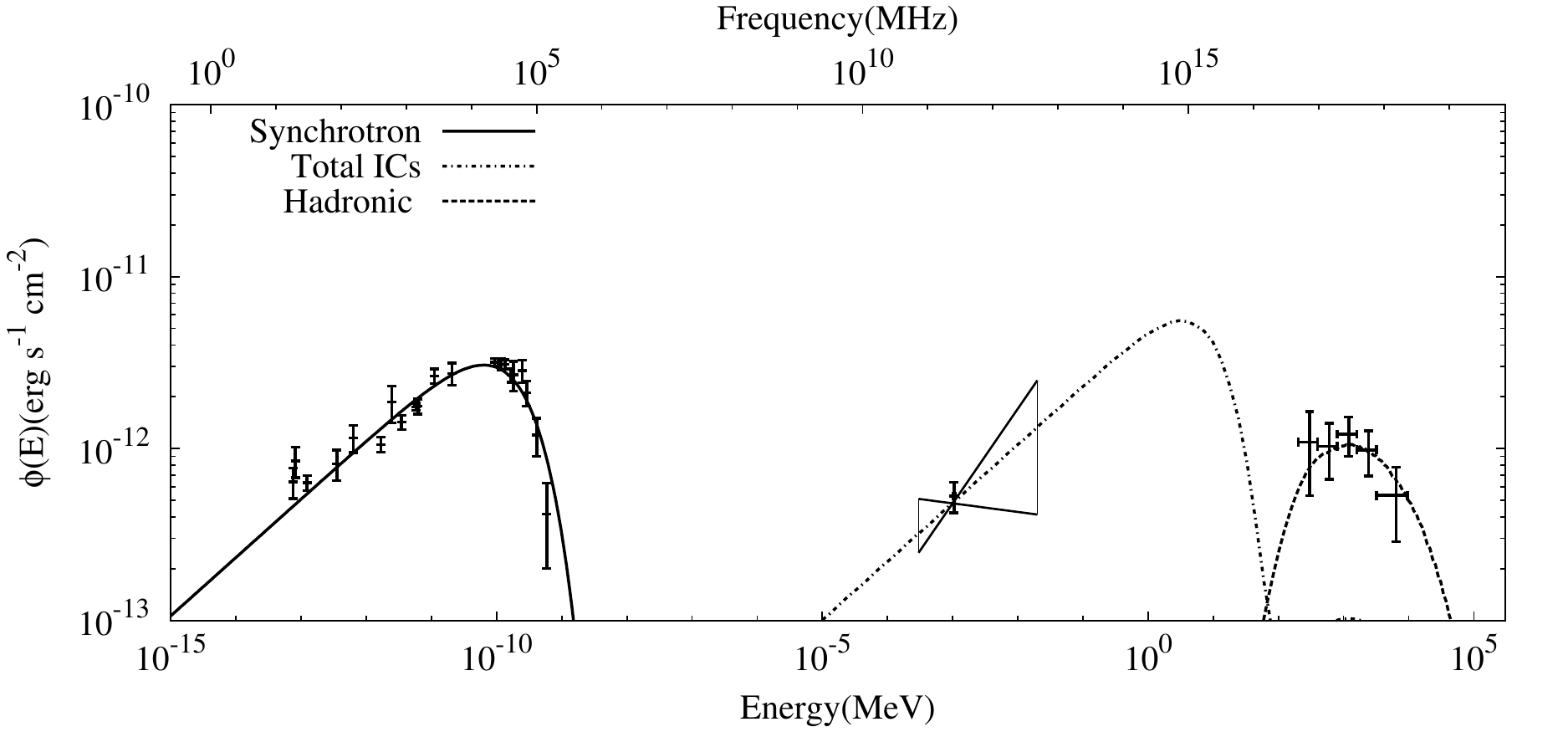}
\caption{The broadband SED of Fornax~A for Model~III, where the X-ray emission is the result of IC scattering by the synchrotron-emitting electrons and the $\gamma$-rays are produce by p-p collisions with pion decay. The model parameters are described in Table~\ref{model_parameters}.}
\label{ForA_had}
\end{figure*}

Model~III assumes that the radio and X-ray flux from the radio lobes result from synchrotron and IC scattering, respectively, while p-p collisions with pion decay account for the $\gamma$-rays. The electron energy distribution is as for Model~I and the proton energy distribution is also described by a simple power law, $N_p(E)=C (E/1{\rm GeV)}^{r} e^{-(E/E_{cp})}$, where $C$ is a normalisation constant, $E$ is energy, $r$ is the proton energy index and $E_{cp}$ is the high cutoff energy. As shown in Fig. \ref{ForA_had}, Model~III fits all the data points and predicts the observed X-ray spectrum. However, when we calculate the total energy budget of protons, assuming a thermal proton density in the lobes of $3\times 10^{-4}~\rm cm^{-3}$ \citep{seta}, we obtain a value of $4.6 \times 10^{60}~\rm erg $, which is two orders of magnitude larger than the thermal emission of the lobes \citep{seta}. This conclusion may be avoided if the emission is localised to relatively denser substructures within the lobes; we explore this scenario briefly in Section~5.

The $\chi^2$ per degree of freedom values, as listed in Table~\ref{model_parameters}, are all greater than unity. The large values are due mainly to some of the radio and microwave data points having small estimated errors, despite their relatively large scatter above and below the smooth models. For Models~I and II there are 22 degrees of freedom (27 independent observations, minus 4 fitted parameters, minus 1) and the $\chi^2$ per degree of freedom values are, 76/22 and 67/22, respectively. These values are significantly higher than the value of 39/19 for model~III, which has fewer degrees of freedom due to an additional 3 fitted parameters relating to the proton energy distribution. Notwithstanding the $\chi^2$ per degree of freedom values, Model~I is ruled out due to the incorrect X-ray flux and Model~II is ruled out based on the incorrect X-ray spectrum. Hence, the model that best fits the data is Model~III.  

\begin{table*}
\centering 
\caption{Summary of Fornax~A SED best-fit model parameters.}
\label{model_parameters}
\begin{tabular}{@{}lccc} 
 \hline
Model components & Model~I & Model~II & Model~III\\
 \hline
Magnetic field strength ($\mu$G)  & $0.87\pm0.05$ & $0.90\pm0.05$ & $2.6\pm0.3$\\
Electron energy index & $-2.32\pm0.04$ & $-2.32\pm0.03$ & $-2.6\pm0.2$\\
Electron high energy cutoff (GeV) & $70\pm4$ & $70\pm4$ & $45\pm5$\\
Electron low energy cutoff (GeV) & N/A & 0.9 (fixed) & N/A\\
Electron normalization ($\times10^{-9}$) & $0.30\pm0.05$ & $0.30\pm0.05$ & $0.05\pm0.01$ \\
Proton energy index & N/A & N/A & $-2.0\pm0.3$\\
Proton high energy cutoff (GeV) & N/A & N/A & $120 \pm 30$\\
$\chi^2$/d.o.f. & 76/22 & 67/22 & 39/19 \\ 
\hline
\end{tabular}
\end{table*}

\section{Discussion}

Our analysis of the low-frequency radio spectrum of the Fornax~A lobes confirms that the assumption implicit in the work of \citet{kaneda}, \citet{isobe} and \citet{tashiro}, that both lobes' spectra can be described by the same power-law, is a valid one. We find, using spectral tomography and spectral-index mapping, that both of the radio lobes have a spectral index of $\alpha_{\rm r} = -0.77 \pm 0.05({\rm stat})^{+0.09_{\rm sys}}_{-0.08_{\rm sys}}$. Taking into account systematic errors, our results are consistent with the radio spectral index of $\alpha_r=-0.68$ used by \citet{isobe} and \citet{tashiro} to model the X-ray IC emission. Our flux-density measurements and spectral indices for the lobes are also consistent with the MWA 32-tile results of \citet{bernardi} at 189~MHz.

Our spectral tomography results reveal a complex central region in Fornax~A, consisting of three overlapping components of differing spectral index; the lobes, the core and the bridge, and our spectral-index mapping results reveal a subtle steepening of the spectrum from the edges of the lobes in towards the core. In this respect, parallels can be drawn between Fornax~A and another well-studied radio galaxy, Cygnus~A, which also has a complex, filamentary lobe structure \citep{carilli} and a bridge connecting the two lobes that steepens in spectral index toward its compact core \citep{carilli,carilli1991,swarup,winter}. Synchrotron ageing models have been used to explain this spectral behaviour, which results from particle acceleration occurring primarily where the relativistic particles interact with the surrounding intergalactic medium and the source expanding out, leaving behind an `older' population of electrons, in which the highest energy particles have been depleted through synchrotron radiative processes \citep{pacholczyk,scheuer}. This process could also be responsible for the bridge spectrum in Fornax~A. 

Bridge structures such as that observed in Fornax~A are common in high-powered radio galaxies, as observed by \citet{leahy}. In most cases, the bridges are distorted \citep{leahy}, as is the case with Fornax~A, where the core and the bridge are offset from each other. \citet{ekers} propose that the offset is due to the active galaxy being perturbed by an infalling galaxy around $10^9$ years ago. In this scenario, the bridge is a remnant of the beamed emission before the merger, and has been left behind by the central galaxy as it moves northward through the intergalactic medium. Our spatially-resolved spectral analysis supports the \citet{ekers} scenario, since the relic bridge emission is shown to have a steep spectrum as a result of radiative losses since the merger occurred.

\citet{lanz} present the discovery of two regions of low X-ray surface brightness regions, centred to the south-west and south-east of the Fornax~A core. They interpret these as lower gas density `X-ray cavities', and suggest that they are likely to be filled with relativistic plasma that is undetected at 1.4~GHz, but should be detected at lower frequencies. Our observations at 154~MHz show that there is diffuse, steep-spectrum emission in these X-ray cavities, supporting the \citet{lanz} scenario for the formation of X-ray cavities close to the nucleus of Fornax~A.

When we attempt to model the radio, microwave, X-ray and $\gamma$-ray data, under the assumption that the same population of synchrotron-emitting electrons is producing the X-ray and $\gamma$-ray emission via IC scattering of CMB and EBL photons, we are unable to find an acceptable fit. Our Model~I over-predicts the observed X-ray flux by an order of magnitude and our Model~II produces an X-ray spectrum that is much harder than the observed values. We therefore reject the hypothesis that both the X-ray and $\gamma$-ray emission in the Fornax~A lobes are purely the result of IC scattering.

The SED data points are fit well by our Model~III, in which the $\gamma$-ray emission is produced by p-p collisions with pion decay. However, the total energy budget of the protons in this model is two orders of magnitude larger than the thermal energy suggested by \citet{seta}, which is physically unrealistic. The excess energy problem may be overcome, however, if the p-p collisions are taking place primarily in the filamentary structures of the lobes identified at 1.4~GHz by \citet{fomalont}. This idea has been proposed as an explanation for the $\gamma$-ray emission observed in the so-called {\it Fermi} bubbles of our own Galaxy by \citet{crocker1} and \citet{crocker2}. Adiabatic compression of the target gas and magnetic fields into filaments, as suggested by \citet{crocker2}, could produce the observed $\gamma$-rays with a total proton energy that is similar to, or less than, the observed thermal energy reported by \citet{seta}, if the filling factor is small. In this scenario, the concentrated magnetic fields in the filaments compensate for the small filling factor in order to produce the observed synchrotron emission. In reality, a combination of IC scattering and p-p collisions both inside and outside of the lobe filaments is probably responsible for the total high-energy emission in Fornax~A. Incorporating this into a fully time-dependent, self-consistent model of the lobes is, however, left for future work.

Fornax~A was identified by \citet{EBL2} as an ideal target for {\it measuring} the EBL, under the assumption that the $\gamma$-ray emission detected by {\it Fermi}-LAT was produced by IC scattering. However, since we have the observational constraints on the EBL from \citet{EBL1}, and we have identified that processes other than IC scattering may be responsible for the $\gamma$-rays, we have not pursued this measurement.

The position of the $\gamma$-ray source that we associate with Fornax~A is consistent with the position of the host galaxy NGC~1316, so we cannot rule out the possibility that the $\gamma$-ray emission is originating in the host galaxy rather than the lobes. However, we consider this unlikely as the core is much weaker than the lobes at both radio (\citealt{ekers,fomalont}, this work) and X-ray \citep{iyomoto} wavelengths. \citet{iyomoto} interpret the weaker core as evidence that the AGN is currently in a state of declined activity. Our spectral-tomography results indicate that the core region has a steeper spectral index than the lobes, with a value of $-0.88 \pm 0.10_{\rm{stat}}$, compared to the lobes with a spectral index of $-0.77 \pm 0.05_{stat}$. This could indicate an older spectral age for the core, due to the period of declined activity, but could also be related to different emission processes and optical depths in the core and in the lobes. A lack of AGN activity does not necessarily preclude a $\gamma$-ray detection, as $\gamma$-ray emission has been confirmed from a number of `normal' galaxies including the Large and Small Magellanic Clouds, M31 and a number of known starburst galaxies \citep{2fgl,1fgl}. However, since NGC~1316 is neither extremely close-by nor undergoing intense star formation, we consider it unlikely that the host galaxy is producing the detected $\gamma$-rays.

\section{Conclusion}

We have presented new low-frequency observations of Fornax~A at 154~MHz from the MWA and used these data, along with previously published data at 1510~MHz, to conduct a spatially resolved study of the spectral index of Fornax~A. We have also presented microwave flux densities obtained from \emph{Planck} and {\it WMAP} data and $\gamma$-ray flux densities from \emph{Fermi}-LAT data and used these, in combination with previously published flux-density measurements at radio and X-ray energies, to model the spectral energy density of Fornax~A. Our results best support a scenario where the X-ray photons are produced by inverse-Compton scattering of the cosmic microwave background and extragalactic background light by the radio-synchrotron emitting electrons in the lobes, while the $\gamma$-rays are the result of proton-proton collisions localised in the lobe filaments.

\section*{Acknowledgments}

This scientific work makes use of the Murchison Radio-astronomy Observatory, operated by CSIRO. We acknowledge the Wajarri Yamatji people as the traditional owners of the Observatory site. Support for the MWA comes from the U.S. National Science Foundation (grants AST-0457585, PHY-0835713, CAREER-0847753, and AST-0908884), the Australian Research Council (LIEF grants LE0775621 and LE0882938), the U.S. Air Force Office of Scientific Research (grant FA9550-0510247), and the Centre for All-sky Astrophysics (an Australian Research Council Centre of Excellence funded by grant CE110001020). Support is also provided by the Smithsonian Astrophysical Observatory, the MIT School of Science, the Raman Research Institute, the Australian National University, and the Victoria University of Wellington (via grant MED-E1799 from the New Zealand Ministry of Economic Development and an IBM Shared University Research Grant). The Australian Federal government provides additional support via the Commonwealth Scientific and Industrial Research Organisation (CSIRO), National Collaborative Research Infrastructure Strategy, Education Investment Fund, and the Australia India Strategic Research Fund, and Astronomy Australia Limited, under contract to Curtin University. We acknowledge the iVEC Petabyte Data Store, the Initiative in Innovative Computing and the CUDA Center for Excellence sponsored by NVIDIA at Harvard University, and the International Centre for Radio Astronomy Research (ICRAR), a Joint Venture of Curtin University and The University of Western Australia, funded by the Western Australian State government. We acknowledge the support of the projects Spanish MINECO AYA2012-39475-C02-01 and CSD2010-00064. We would also like to thank the referee for their useful comments and suggestions.

\bsp

\label{lastpage}

\end{document}